\documentclass[namedreferences]{SolarPhysics}
\usepackage[optionalrh]{spr-sola-addons} % For Solar Physics
\usepackage{graphicx}                    % For eps figures, newer & more powerful
\usepackage{bm}
\usepackage{color}                       % For color text: \color command
\usepackage{url}                         % For breaking URLs easily trough lines
\begin{document}

\begin{article}

\begin{opening}

\title{Cross Helicity and Turbulent Magnetic Diffusivity in the Solar Convection Zone}

%%%%%%%%%%%%%%%%%%%%%%%%%%%%%%%%%%%%%%%%%%%%%%%%%%%
%% Authors Names
%
\author{G.~R\"udiger{}$^{1}$\sep L.L.~Kitchatinov{}$^{2,3}$\sep  A.~Brandenburg$^{4,5}$}

%%%%%%%%%%%%%%%%%%%%%%%%%%%%%%%%%%%%%%%%%%%%%%%%%%%
%% Runningheads
%
\runningauthor{G. R\"udiger {\it et al.}}
\runningtitle{Magnetic cross helicity}

%%%%%%%%%%%%%%%%%%%%%%%%%%%%%%%%%%%%%%%%%%%%%%%%%%%
%% Affiliations
%
  \institute{$^{1}$ Astrophysikalisches Institut Potsdam, An der Sternwarte 16, D-14482 Potsdam, Germany,
                     email: gruediger@aip.de \\
   $^{2}$ Institute for Solar-Terrestrial Physics, P.O. Box 291, Irkutsk 664033, Russia, email: kit@iszf.irk.ru \\
$^{3}$ Pulkovo Astronomical Observatory, St. Petersburg 196140, Russia\\
$^{4}$ Nordita, AlbaNova University Center, Roslagstullsbacken 23, SE-10691 Stockholm, Sweden\\
$^{5}$ Department of Astronomy, Stockholm University, SE-10691 Stockholm, Sweden
 }

%%%%%%%%%%%%%%%%%%%%%%%%%%%%%%%%%%%%%%%%%%%%%%%%%%%

\begin{abstract}
In a density-stratified turbulent medium the cross helicity $\langle
\mathrm{\bf u}'\cdot \mathrm{\bf B}'\rangle$ is considered as a
result of the interaction of the velocity fluctuations and a
large-scale magnetic field. By means of a quasilinear theory and by
numerical simulations we find the cross helicity and the mean
vertical magnetic field anti-correlated. In the high-conductivity
limit the ratio of the helicity and the mean magnetic field equals
the ratio of the magnetic eddy diffusivity and the (known) density
scale height. The result can be used to predict that the cross
helicity at the solar surface exceeds the value of $1$
gauss\,km\,s$^{-1}$. Its sign is anti-correlated with that of the
radial mean magnetic field. Alternatively, we can use our result to
determine the value of the turbulent magnetic diffusivity from
observations of the cross helicity.
\end{abstract}

%%%%%%%%%%%%%%%%%%%%%%%%%%%%%%%%%%%%%%%%%%%%%%%%%%%
%% Keywords
%

\end{opening}
%-------------------------------------------------

%%%%%%%%%%%%%%%%%%%%%%%%%%%%%%%%%%%%%%%%%%%%%%%%%%%
%% Sections
%
\section{Introduction}
%%%%%%%%%%%%%%%%%%%%%%%%%%%%%%%%%%%%%%%%%%%%%%%%%%%%%%%%%%%%%%%%%%%%%%%%%%%%%%%
Dynamo theory for convective zones needs to know both the values of
the $\alpha$-effect and the eddy diffusivity. The $\alpha$-effect is
strongly related to the kinetic helicity which has opposite signs in
the two hemispheres. Almost all of the theoretical calculations for
rotating stratified turbulence lead to negative helicity ({\it i.e.}
positive $\alpha$-effect) for the northern hemisphere and positive
helicity ({\it i.e.} negative $\alpha$-effect) for the southern
hemisphere. Despite all of the complications to measure the helicity
on the solar surface, a new result has recently been presented by
Komm, Hill, and Howe (2008). They do indeed find negative (positive)
values for the kinetic helicity in the northern (southern)
hemisphere. This result is based on a ring-diagram analysis of GONG
data. The values remained constant as long as the magnetic field did
not exceed 10\,gauss (G). Using the observation of the DIV-CURL
correlation, ${\cal C}=\langle
(u_{,x}+v_{,y})(v_{,x}-u_{,y})\rangle$, which is proportional to the
kinetic helicity (R\"udiger, Brandenburg, and Pipin, 1999; see
R\"udiger and Hollerbach (2004) for more details), Duvall and Gizon
(2000) found $\cal C$ to be negative (positive) in the northern
(southern) hemisphere, as derived from the horizontal velocity
components of mesogranulation patterns. Egorov, R\"udiger, and
Ziegler (2004) simulated these observations with the NIRVANA code
and reproduced them with Taylor numbers as small as $10^3$.

By use of the finding of Keinigs (1983) the current helicity and the
$\alpha$-effect are anti-correlated, so one also can derive the sign
of the $\alpha$-effect by observation of the current helicity
$\langle \mathrm{\bf J}'\cdot  \mathrm{\bf B}'\rangle$. Seehafer
(1990) started to observe the current helicity at the solar surface
showing that it is negative (positive) in the northern (southern)
hemisphere. Again the $\alpha$-effect is found to be positive
(negative) in the northern (southern) hemisphere.

The numerical value of the helicity derived by Komm {\it et al.}
(2008) is of the order of $10^{-7}$ cm\,s$^{-2}$. This turns out to
be very small, because the resulting $\alpha$-effect is below 1
cm\,s$^{-1}$. By comparison, K\"apyl\"a, Korpi, and Brandenburg
(2009) find $\alpha \simeq 0.03 u_{\rm rms}$ near the surface from
their convection simulations. With $u_{\rm rms}\simeq 300$
m\,s$^{-1}$, this corresponds to 10 m\,s$^{-1}$. The maximal
$\alpha$-value in their box center is of the order of 0.3$u_{\rm
rms}$. This highlights a major discrepancy between theory and
observations or, at least, a difficulty in determining $\alpha$ from
observations.

The empirical definition of the turbulent magnetic diffusion seems
to be more straightforward. The decay of non-permanent magnetic
structures such as sunspots or larger active regions lead to
numerical values of the turbulent magnetic diffusivity. One finds
$\eta_{\rm T}\simeq 10^{11}$ cm$^2$\,s$^{-1}$ from sunspot decay
(Stix, 1989) or $\eta_{\rm T}\simeq 10^{12}$ cm$^2$\,s$^{-1}$ from
the decay of active regions (Schrijver and Martin, 1990). These
values are smaller than the value of $3\times 10^{12}$
cm$^2$\,s$^{-1}$, which results from the widely used formula
$\eta_{\rm T}\approx u_{\rm rms} \ell /3$ with correlation or mixing
length $\ell$ and parameter values taken close to the surface. There
is no possibility until now to observe the turbulent diffusivity on
the solar surface for the quiet Sun where the magnetic quenching of
this quantity by magnetic fields is negligible. We shall demonstrate
in the present paper that there is a rather simple possibility to
observe the magnetic diffusivity even in the presence of very weak
magnetic fields ($< 10$\,G), for which quenching should be
negligible.

We shall show that in the presence of a mean magnetic field along
the direction of density stratification, hydromagnetic turbulence
will attain cross helicity, $\langle \mathrm{\bf u}'\cdot
\mathrm{\bf B}'\rangle$, whose value is proportional to the
turbulent magnetic diffusivity. Indeed, our work is an extension of
that by Kleeorin {\it et al.} (2003), who considered the effect of
stratification of turbulent intensity. More recently, Kuzanyan,
Pipin, and Zhang (2007) emphasized the importance of cross-helicity
for estimating internal solar parameters important for the dynamo.
We mention in this connection that cross helicity itself may
constitute a potentially important dynamo effect \cite{Yos90,Yok96}.
In this paper we propose a measurement of cross-helicity in the Sun
for estimating the turbulent magnetic diffusivity in quiet regions.
We argue that this is more accurate than measuring, for example, the
mean electromotive force.

%%%%%%%%%%%%%%%%%%%%%%%%%%%%%%%%%%%%%%%%%%%%%%%%%%%%%%%%%%%%%%%%%%%%%%%%%%%%%%%%
 \section{Mean-Field Electrodynamics}\label{sect1}
%%%%%%%%%%%%%%%%%%%%%%%%%%%%%%%%%%%%%%%%%%%%%%%%%%%%%%%%%%%%%%%%%%%%%%%%%%%%%%%%%
Let $\mathrm{\bf u}' = \mathrm{\bf u} - \langle \mathrm{\bf
u}\rangle$ and $\mathrm{\bf B}' = \mathrm{\bf B} -
\langle\mathrm{\bf B}\rangle$ be the fluctuations of velocity and
magnetic field about an average value denoted by angular brackets.
The mean-field dynamo theory of cosmic magnetic fields is based on
the relation
\begin{equation}
\langle \mathrm{\bf u}'\times \mathrm{\bf B}'\rangle= \alpha \langle
\mathrm{\bf B}\rangle - \beta \langle \mathrm{\bf J} \rangle
\label{eq1}
\end{equation}
between the turbulent electromotive  force ${\bf {\cal E}}= \langle
\mathrm{\bf u}'\times \mathrm{\bf B}'\rangle$ and the mean-field
quantities $\langle \mathrm{\bf B}\rangle$ and $\langle \mathrm{\bf
J}\rangle$, where $\mathrm{\bf J}$ is the mean current density. For
the purpose of this discussion we neglect here possible effects of
mean flows on the correlators, {\it i.e.} we assume $\langle
\mathrm{\bf u}\rangle=0$. Note the basic difference between the
quantities $\alpha$ and $\beta$ in that $\beta$  is a scalar while
$\alpha$ is a  pseudoscalar. For rotating stars a pseudoscalar can
be formed by use of the basic rotation rate, {\it e.g.}, $\alpha
\propto (\mathrm{\bf g}\cdot{\bf\Omega})$ with gravity $\mathrm{\bf
g}$ as the only remaining preferred direction apart from
$\bf{\Omega}$. Hence, the amplitude of the $\alpha$-effect must
mainly be influenced by the Coriolis number
\begin{equation}
\Omega^*=2\tau_{\rm corr} \Omega,
\label{corrnum}
\end{equation}
where $\tau_{\rm corr}$ is the correlation time of the dominating mode
of turbulence.
However, $\Omega^*$ is very small at the solar surface so that the
$\alpha$-effect in Equation (\ref{eq1}) cannot be observed directly.

The parameter $\beta$ in Equation (\ref{eq1}) exists even in
nonrotating plasmas. It is thus not governed by the Coriolis number
$\Omega^*$ and is therefore not small by comparison. It is, however,
not possible to observe by direct means the mean current density
$\langle \mathrm{\bf J}\rangle$ at the solar surface. The decay of
sunspots should provide  good estimates for $\beta$ when the
induction equation is solved by using Equation (\ref{eq1}); and the
time-dependent solutions are compared with the observations (Krause
and R\"udiger, 1975). Though successful, this procedure cannot serve
as a proof of the existence of Equation (\ref{eq1}).
%AB: what is mean with the sentence above?
We must conclude, therefore, that the basic Equation (\ref{eq1}) cannot
be tested with observations  taken from the solar surface.
This is an unsatisfying situation given that Equation (\ref{eq1}) is a
fundamental relation of a whole branch of cosmic MHD and, of course,
there is no better laboratory than the Sun to probe such basic relations.

Fortunately, the situation is quite different for another
correlation between fluctuations of flow and field, namely the cross
helicity $\langle \mathrm{\bf u}'\cdot \mathrm{\bf B}'\rangle$,
which itself is a pseudoscalar. It is straightforward to formulate
the relation
\begin{equation}
\langle \mathrm{\bf u}'\cdot \mathrm{\bf B}'\rangle=
\alpha_\mathrm{c} \langle \mathrm{\bf g}\cdot \mathrm{\bf B} \rangle
- \beta_\mathrm{c}\langle \bf{\Omega}\cdot \mathrm{\bf J}\rangle
\label{eq2}
\end{equation}
similar to Equation (\ref{eq1}). The $\alpha_\mathrm{c}$-effect does
{\em not} run with the Coriolis number $\Omega^*$. Similar to the
$\alpha$-effect in Equation (\ref{eq1}) the $\alpha_\mathrm{c}$ in
Equation (\ref{eq2}) is of the dimension of a velocity but this
velocity should be much larger than the corresponding $\alpha$ in
Equation (\ref{eq1}). As the second term on the RHS of Equation
(\ref{eq2}) only exists in the presence of rotation, it will be
negligibly small at the solar surface.

In summary, by simple reasons the observations of the cross
correlation $\langle \mathrm{\bf u}'\cdot \mathrm{\bf B}'\rangle$ at
the solar surface should give a realistic  chance to confirm the
existence of relations that are typical for mean-field
electrodynamics.

%%%%%%%%%%%%%%%%%%%%%%%%%%%%%%%%%%%%%%%%%%%%%%%%%%%%%%%%%%%%%%%%%%%%
\section{Nonconservation of Cross-Helicity in Turbulent Fluids}
%%%%%%%%%%%%%%%%%%%%%%%%%%%%%%%%%%%%%%%%%%%%%%%%%%%%%%%%%%%%%%%%%%%%
The cross-helicity is conserved globally (as volume integral) in
ideal incompressible fluids \cite{Wol58}. In view of the conservation law, it may
be anticipated that the balance of small-scale cross-helicity should
be treated globally by defining the small-scale cross-helicity
sources, cross-helicity fluxes, and formulating the dynamical
cross-helicity equation similar to the approach used to study the
balance of magnetic helicity. Examples of such an approach to the
cross-helicity problem can be found in the literature \cite{SB09}.
In this section we show, however, that cross-helicity is not conserved in
turbulent fluids such as the solar convection zone and its balance
is controlled by local processes.

The turbulence is known to dissipate efficiently the quantities,
which are conserved in ideal fluids (with zero diffusivities). The
well known example is the energy balance. Kinetic energy is
conserved in ideal hydrodynamics. The rate of energy dissipation in
Kolmogorov (1941) picture of turbulence is, however, constant
independent of whatever small (but finite) is the viscosity. The
same is true about almost all quantities conserved in ideal fluids.
Turbulent fragmentation of scales cascades rapidly the quantities to
the smallest scales where they dissipate.

The only known exception is magnetic helicity that is conserved
even in turbulent fluids. The reason can be seen from the following.
In the simplest case of isotropic homogeneous turbulence, the
spectrum tensor of fluctuating magnetic fields can be written as
\begin{equation}
    B_{ij}(\mathrm{\bf k}) = \frac{E^\mathrm{m}(k)}{8\pi k^2}
    \left(\delta_{ij} - \frac{k_ik_j}{k^2}\right) -
    \frac{\mathrm{i}H^\mathrm{m}(k)}{8\pi k^2}\varepsilon_{ijn}k_n ,
    \label{spect}
\end{equation}
where $E^\mathrm{m}$ and $H^\mathrm{m}$ are magnetic energy
and helicity spectra,
\begin{equation}
  \langle B'^2\rangle = \int\limits_0^\infty E^\mathrm{m}(k)\mathrm{d}k,\ \ \
  \langle\mathrm{\bf  B}'\cdot\mathrm{\bf  A}'\rangle = \int\limits_0^\infty
  H^\mathrm{m}(k) \mathrm{d}k,\ \ \ \mathrm{\bf B}' = \mathrm{rot}\mathrm{\bf  A}' .
  \label{int}
\end{equation}
The spectrum tensor (\ref{spect}) is positive definite,
$B_{ij}C_iC^*_j \geq 0$ \cite{B33}, where $\mathrm{\bf C}$ is an
arbitrary vector and the asterisk marks complex conjugation. For the
tensor (\ref{spect}), this leads to the inequality \cite{Mof69}
\begin{equation}
  | H^\mathrm{m}(k)| k \leq E^\mathrm{m}(k) ,
  \label{uneq}
\end{equation}
which is also known as the realizability condition. Imagine that at
some (small) wavenumber $k_1$ helicity is finite and the spectral
magnetic helicity to energy ratio at that wavenumber is $k_0^{-1} =
|H^\mathrm{m}(k_1)|/E^\mathrm{m}(k_1)$. If helicity could follow
magnetic energy in its cascade to large $k$, the ratio
$H^\mathrm{m}/E^\mathrm{m}$ would be constant across the spectrum.
Then, Equation (\ref{uneq}) would require $k\leq k_0$, which is an
inequality that is impossible to satisfy for spectra with
sufficiently broad inertial range. Therefore, magnetic helicity
cannot be cascaded to diffusive scales to dissipate. The helicity is
conserved, and the conservation law has important consequences for
large-scale dynamos \cite{BS05}.

The conservation of magnetic helicity is, however, an exception.
For example, for {\it kinetic} helicity, which is also conserved in
ideal hydrodynamics, instead of inequality (\ref{uneq}) we have
$|H^\mathrm{k}|\leq E^\mathrm{k} k$ with no restrictions for the kinetic
helicity cascade to viscous scales.
As a consequence, kinetic helicity is not conserved.
The same is true for the cross helicity.
This can also be seen by comparing the rates of helicity dissipation.
Using the fact that vorticity and current density scale inversely proportional
to the square roots of viscosity and magnetic diffusivity, respectively,
we see that the rate of magnetic helicity dissipation decreases with decreasing
magnetic diffusivity proportional to its square root, while that of cross helicity
is independent of viscosity and magnetic diffusivity and does therefore not vanish.
The cross-helicity balance is controlled by local processes.
In spite of some striking similarities in the saturation of dynamos
controlled by magnetic and cross helicity, the presence of significant
cross-helicity dissipation as well as the forcing term in the momentum
equation destroy the nice analogy \cite{SB09}.
In the following we proceed with deriving the cross-helicity from local
relations.

%%%%%%%%%%%%%%%%%%%%%%%%%%%%%%%%%%%%%%%%%%%%%%%%%%%%%%%%%%%%%%%%%%%%
\section{Quasilinear Theory of Cross Helicity}\label{derivations}
%%%%%%%%%%%%%%%%%%%%%%%%%%%%%%%%%%%%%%%%%%%%%%%%%%%%%%%%%%%%%%%%%%%%
In this section we derive the symmetric part $\langle
u'_iB'_j\rangle^\mathrm{s} =$ $(\langle u'_iB'_j\rangle + \langle
u'_jB'_i\rangle)/2$ of the cross correlation tensor $\langle
u'_iB'_j\rangle$. The pseudotensor $\langle u'_iB'_j\rangle$ can be
finite only in the presence of a mean magnetic field
$\langle\mathrm{\bf B}\rangle$ and for inhomogeneous fluids. The
required inhomogeneity can be due to stratification of density or
turbulent intensity as well as the inhomogeneity of the mean field
itself.

The turbulent flow is assumed anelastic, so that
$\mathrm{div}(\rho\mathrm{\bf u}') = 0$. It is convenient to use the
Fourier transformation of the momentum density $\mathrm{\bf m} =
\rho\mathrm{\bf u}'$, {\it i.e.}
\begin{equation}
    \mathrm{\bf m}(\mathrm{\bf r},t) = \int\hat{\mathrm{\bf m}}(\mathrm{\bf k},\omega)\
        \mathrm{e}^{\mathrm{i}(\mathrm{\bf k}\cdot\mathrm{\bf r}
    - \omega t)}\mathrm{d}\mathrm{\bf k}\ \mathrm{d}\omega ,
    \label{1}
%Fourier transform
\end{equation}
and similarly for the fluctuation of the magnetic field.
The linearized equation for magnetic fluctuations in terms of the Fourier
amplitudes reads
\begin{eqnarray}
    (-\mathrm{i}\omega &+&\eta k^2)\hat{B}'_i(\mathrm{\bf k},\omega) =
        \nonumber \\
    &=& \mathrm{i} k_j\int\left(\hat{m}_i(\mathrm{\bf k}-\mathrm{\bf k}',\omega-\omega')
    \hat{\left(\frac{B_j}{\rho}\right)}(\mathrm{\bf k}',\omega')-\right.
    \nonumber \\
    &-&\left. \hat{m}_j(\mathrm{\bf k}-\mathrm{\bf k}',\omega-\omega')
    \hat{\left(\frac{B_i}{\rho}\right)}(\mathrm{\bf k}',\omega')\right)\
    \mathrm{d}\mathrm{\bf k}'\mathrm{d}\omega ' ,
    \label{2}
\end{eqnarray}
where $\hat{\mathrm{\bf B}}$ is the Fourier transform of the mean
magnetic field.

The spectral tensor of the momentum density that accounts for the stratification of the turbulence to first order terms reads
\begin{eqnarray}
    &&\langle\hat{m}_i (\mathrm{\bf z},\omega )\hat{m}_j(\mathrm{\bf z}',\omega ')\rangle =
    \delta (\omega + \omega ') \frac{\hat{q}(k,\omega ,\mathrm{\bf \kappa})}{16\pi k^2}\ \times
    \nonumber \\
    &&\quad\quad\quad\quad\quad\quad \times\left(\delta_{ij} -k_ik_j/k^2
    + \left(\kappa_i k_j - \kappa_j k_i\right) /(2k^2)\right),
    \label{3}
\end{eqnarray}
where $\mathrm{\bf k} = (\mathrm{\bf z}-\mathrm{\bf z}')/2,\
\bm{\kappa} = \mathrm{\bf z} + \mathrm{\bf z}'$, $\hat{q}$ is the
Fourier transform of the local spectrum,
\begin{equation}
    q(k,\omega,{\mathrm{\bf r}}) = \rho^2E(k,\omega ,\mathrm{\bf r}) =
        \int \hat{q}(k,\omega , \bm{\kappa})
    \mathrm{e}^{ \mathrm{i}\bm{\kappa}\cdot\mathrm{\bf r}}\ \mathrm{d}\bm{\kappa},
    \label{4}
\end{equation}
so that
\begin{equation}
    \langle u'^2\rangle =
    \int\limits_0^\infty\int\limits_0^\infty E(k,\omega ,\mathrm{\bf r})\
    \mathrm{d}k\,\mathrm{d}\omega .
    \label{5}
\end{equation}
Derivation of the cross correlation yields
\begin{eqnarray}
\langle u'_iB'_j\rangle^\mathrm{s} &=& \frac{1}{2}\eta_\mathrm{T}
    \left(G_i \langle B_j\rangle + G_j \langle B_i\rangle \right)
        + \left(\frac{1}{10}\eta_\mathrm{T} + \frac{4}{15}\hat{\eta}\right)
    \delta_{ij}\left(\mathrm{\bf U}\cdot\langle \mathrm{\bf B}\rangle \right)+
    \nonumber \\
    &+&\left(\frac{1}{10}\eta_\mathrm{T} - \frac{1}{15}\hat{\eta}\right)
    \left(U_i \langle B_j\rangle + U_j \langle B_i\rangle \right)
    \nonumber \\
    &-&\left( \frac{3}{10}\eta_\mathrm{T} + \frac{2}{15}\hat{\eta}\right)
    \left(\langle B_{j,i}\rangle + \langle B_{i,j}\rangle \right) ,
    \label{6}
\end{eqnarray}
where $\mathrm{\bf G} = \mathrm{\bf \nabla}\mathrm{log}\rho $ and
$\mathrm{\bf U} = \mathrm{\bf \nabla}\mathrm{log} \langle
u'^2\rangle$ are the gradients of density and turbulent intensity
and
\begin{eqnarray}
   \eta_\mathrm{T} &=& \frac{1}{3}\int\limits_0^\infty\int\limits_0^\infty
    \frac{\eta k^2 E}{\omega^2+\eta^2k^4}\mathrm{d}k\,\mathrm{d}\omega ,\label{7.1}\\
    \hat\eta &=& \int\limits_0^\infty\int\limits_0^\infty
    \frac{\eta k^2\omega^2 E}{(\omega^2+\eta^2k^4)^2}
    \mathrm{d}k\,\mathrm{d}\omega .
\end{eqnarray}
Here $\eta=1/\mu_0 \sigma$ is the molecular magnetic diffusivity.
Both quantities run with the magnetic Reynolds number for low
conductivity ($\sigma \to 0$) and become finite for high
conductivity ($\eta\to 0$). From the cross correlation tensor
(\ref{6})  the cross helicity
\begin{equation}
    \langle\mathrm{\bf u}'\cdot \mathrm{\bf B}'\rangle =
    \eta_\mathrm{T}\left(\mathrm{\bf G}\cdot \langle \mathrm{\bf B}\rangle \right) +
    \left(\frac{\eta_\mathrm{T}}{2} + \frac{2\hat{\eta}}{3}\right)
    \left(\mathrm{\bf U}\cdot\langle \mathrm{\bf B}\rangle \right)
    \label{8}
\end{equation}
is obtained.

Current observations only supply the correlation $\langle u'_rB'_r\rangle$. From Equation~(\ref{6}) we find
\begin{equation}
    \langle u'_rB'_r\rangle = \eta_\mathrm{T} G \langle B_r\rangle
        -\left(\frac{3\eta_\mathrm{T}}{10} + \frac{2\hat{\eta}}{15}\right)
    \left(2\frac{\partial \langle B_r\rangle}{\partial r} - U \langle B_r\rangle\right) ,
    \label{9}
\end{equation}
where $G=G_r$ and $U=U_r$ are the only non-zero radial components of
the stratification vectors.
Further simplifications can be obtained by using the mixing-length
approximation for the turbulence spectrum,
\begin{equation}
    E(k,\omega ,\mathrm{\bf r}) = 2\langle u'^2\rangle
    \delta\left(k - \ell^{-1}\right)\delta (\omega ), \ \ \
    \eta = \ell^2/\tau_{\rm corr}
    \label{10}
\end{equation}
(Kitchatinov, 1991), where $\ell$ is mixing length and $\tau_{\rm
corr}$ is the correlation time. It yields
\begin{equation}
    \langle u'_rB'_r\rangle = \eta_\mathrm{T}
    \left( G \langle B_r\rangle - \frac{3}{5}\frac{\partial \langle B_r\rangle}{\partial r}
    + \frac{3}{10}U\langle B_r\rangle\right).
    \label{11}
\end{equation}
The result can be explained as follows. A rising fluid element $u'_r
> 0$ expands so that $B'_r$ has the opposite sign as $\langle
B_r\rangle$. The fluid particles which go down, $u'_r <0$, compress
and $B'_r$ has the same sign as $B_r$. The sign of the product
$u'_rB'_r$ is opposite to $\langle B_r\rangle$ in both cases -- in
accord with the first term on the right hand side (RHS) of Equation
(\ref{11}); note the negativity of  $G$. An upward divergence of the
mean field reduces the effect of density stratification. This is
realized  by the second term on the RHS of Equation~(\ref{11}). The
third term shows that also the non-uniformity of the turbulent
intensity makes a contribution. However, the contribution of density
stratification is dominant, because the density gradient in the
upper convection zone is larger than the turbulent intensity
gradient. This is already clear from Figure~1 of Krivodubskii and
Schultz (1993), who plot, for a solar structure model, the relative
contributions from $\mathrm{\bf G}$ and $\mathrm{\bf U}$ in the
expression for the usual $\alpha$ effect, where both enter in equal
amounts. Here, however, $\mathrm{\bf U}$ enters with a 3/10 factor
and is even more subdominant. Thus, we conclude that a finite cross
correlation (\ref{11}) indicates the presence of a large-scale
radial field of the opposite sign.

The leading term on the RHS of Equation (\ref{11}) is due to the density
gradient.
The resulting relation then reads
\begin{equation}
\frac{\langle u'_rB'_r\rangle}{\langle B_r\rangle}  = - \frac{\eta_{\rm T}}{H_\rho}.
\label{12}
\end{equation}
The magnetic eddy diffusivity can thus be determined if the LHS of
Equation (\ref{12}) is observed and the density scale height
$H_\rho$ is known from models of the solar atmosphere.
%%%%%%%%%%%%%%%%%%%%%%%%%%%%%%%%%%%%%%%%%%%%%%%%%%%%
\section{Numerical Simulation}
%%%%%%%%%%%%%%%%%%%%%%%%%%%%%%%%%%%%%%%%%%%%%%%%%%%%
It is straightforward to verify the validity of Equation~(\ref{12})
using numerical simulations of isothermally stratified forced
turbulence in a layer with constant gravity, $\mathrm{\bf
g}=(0,0,-g)$ in Cartesian coordinates. In that case the scale
height, $H_\rho=c_{\rm s}^2/g$, is constant.

We perform simulations in a cubic domain of size $L^3$,
so the minimal wavenumber is $k\equiv k_1=2\pi/L$.
We solve the governing equations of compressible magnetohydrodynamics
with an isothermal equation of state.
The flow is driven by a random forcing function consisting of non-helical
waves with wavenumbers whose modulus lies in a narrow band around an
average wavenumber $k_{\rm f}$ (which corresponds to $\ell^{-1}$
used in the previous section).
We arrange the amplitude of the forcing function such that the RMS Mach
number is around 0.1 or less, so the effects of compressibility are
negligible.

In all our runs we adopt stress-free pseudo-vacuum boundary
conditions on the top and bottom boundaries, {\it i.e.} the
horizontal magnetic field vanishes. The magnetic field is expressed
in terms of the vector potential $\mathrm{\bf A}$ as $\mathrm{\bf
B}=\mathrm{\bf B}_0+\mathrm{\bf \nabla}\times\mathrm{\bf A}$, where
$\mathrm{\bf B}_0=(0,0,B_{0z})={\rm const}$ is the imposed vertical
field which is fixed for each run.
The simulations were performed with the {\sc Pencil Code}%
\footnote{{\tt http://pencil-code.googlecode.com}},
which uses sixth-order explicit finite differences in space and
third-order accurate time stepping method (Brandenburg and Dobler, 2002).
A numerical resolution of up to $256^3$ meshpoints was used, depending
on the value of the magnetic Reynolds number.

 \begin{figure}[t!]
 \begin{center}
 \includegraphics[width=\columnwidth]{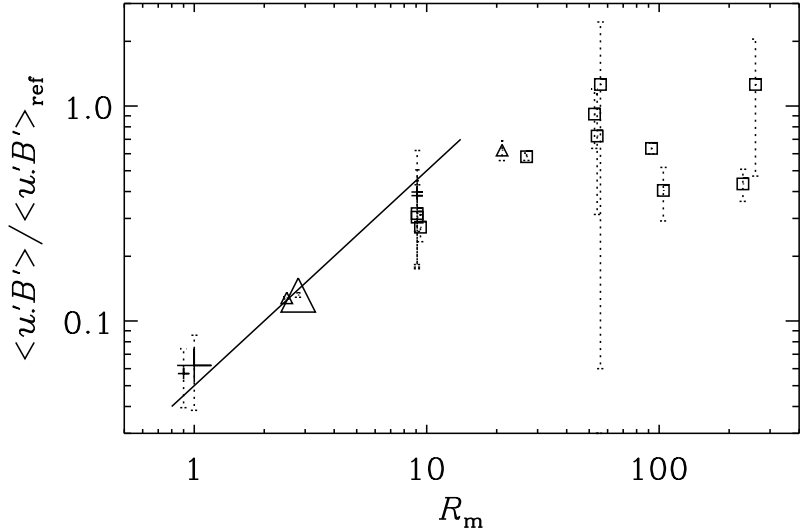}
 \end{center}
 \caption[]{
Dependence of the normalized cross helicity on $R_{\mathrm m}$ for
various field strength $B_z/B_{\rm eq}<0.1$, $P_{\mathrm m}=1$,
$k_{\rm f}/k_1=2.2$, and $H_\rho k_1=2.5$. The straight line denotes
the fit $\langle\mathrm{\bf u}'\cdot\mathrm{\bf B}'\rangle/
\langle\mathrm{\bf u}'\cdot\mathrm{\bf B}'\rangle_{\rm
ref}=0.05R_{\rm m}$.
 }
 \label{pRm_dep}
 \end{figure}

 \begin{figure}[t!]
 \begin{center}
 \includegraphics[width=\columnwidth]{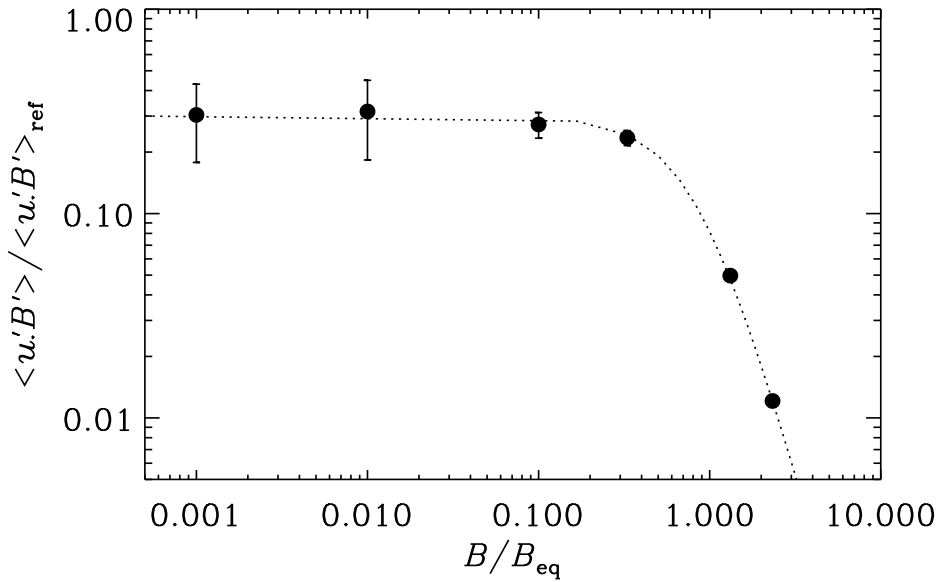}
 \end{center}
 \caption[]{
Dependence of the normalized cross helicity on the normalized
vertical field strength for $R_{\rm m}=10$, $P_{\rm m}=1$, $k_{\rm
f}/k_1=2.2$, and $H_\rho k_1=2.5$. The dotted line corresponds to
the graph of $0.3/[1+(\langle\mathrm{\bf B}\rangle/B_{\rm
ref})^2]^{3/2}$ with $B_{\rm ref}=0.85\,B_{\rm eq}$.
 }
 \label{pB_dep}
 \end{figure}

We perform simulations for a number of different parameter combinations.
The parameters that are being varied include the strength of the imposed
vertical field $B_z$, the forcing wavenumber $k_{\rm f}$,
the gravitational acceleration $g$, and hence $H_\rho$, and the
values of the magnetic diffusivity.
We express these quantities in non-dimensional form and define the
magnetic Reynolds number as
\begin{equation}
R_\mathrm{m}=\frac{u_{\rm rms}}{\eta k_{\rm f}}. \label{Rm}
\end{equation}
The strength of the magnetic field is characterized by the mean
equipartition field strength,
\begin{equation}
B_{\rm eq}=\sqrt{\mu_0\langle\rho\rangle}\,u_{\rm rms},
\label{Beq}
\end{equation}
which is of the order of 1000\,G at the solar surface. We determine
the cross helicity, $\langle\mathrm{\bf u}'\cdot\mathrm{\bf
B}'\rangle$, as a volume average. In order to relate this to
Equation~(\ref{12}) we also need to estimate the value of the
turbulent magnetic diffusivity. Earlier work showed that, to a good
approximation, $\eta_{\rm T}$ can be estimated by Sur, Brandenburg,
and Subramanian (2008),
\begin{equation}
\eta_{\rm T}\approx\eta_{\rm T0}\equiv u_{\rm rms}/3k_{\rm f},
\label{etaT}
\end{equation}
provided $R_{\rm m}\gg1$, {\it i.e.} in the high-conductivity
approximation. In a number of cases we have verified the validity of
this approximation also for the stratified runs shown here.

We present the value of $\langle\mathrm{\bf u}'\cdot\mathrm{\bf
B}'\rangle$ in non-dimensional form by dividing by a reference value
defined after Equation (\ref{12}) as
\begin{equation}
\langle\mathrm{\bf u}'\cdot\mathrm{\bf B}'\rangle_{\rm
ref}=-\frac{\eta_{\rm T0}B_0}{H_\rho}.
\end{equation}

For small $R_{\rm  m}$ the normalized cross helicity depends on
$R_{\rm m}$ (see Figure~\ref{pRm_dep}) but it  reaches unity for
large $R_{\rm m}$. It is the expected behavior as Equation
(\ref{etaT}) gives only a good approximation for Equation
(\ref{7.1}) for the case of high conductivity, {\it i.e.} for
$\sigma \to \infty$. In the opposite case of $\sigma\to 0$,
expression (\ref{7.1}) vanishes so that the small numbers of the
lower-left corner of Figure~\ref{pRm_dep}, become understandable.
For small values of $R_{\rm m}$ we have $\eta_{\rm T} \propto R_{\rm
m}$. For the largest values of $R_{\rm m}$ the error bars for the
numerical results are larger. This is mainly because those
simulations require larger numerical resolution and long run times
become prohibitive.

Figure \ref{pB_dep} shows the dependence of the normalized cross
helicity, defined by the ratio $\langle\mathrm{\bf
u}'\cdot\mathrm{\bf B}'\rangle/ \langle\mathrm{\bf
u}'\cdot\mathrm{\bf B}'\rangle_{\rm ref}$, on the normalized field
strength $B_{0z}/B_{\rm eq}$. Note that the cross helicity is
quenched by nearly a factor of 10 for $B_{0z}\approx B_{\rm eq}$.

%%%%%%%%%%%%%%%%%%%%%%%%%%%%%%%%%%%%%%%%%%%%%%%%%%%%%%%%%
\section{Conclusions}
%%%%%%%%%%%%%%%%%%
We  have shown that nonrotating turbulence at the top of the solar
convection zone under the influence of a vertical magnetic field
yields a finite cross helicity. The only requirement is the
existence of density stratification which enters the induction
equation via the anelastic relation $\mathrm{div}(\rho\mathrm{\bf
u}') = 0$. The Boussinesq approximation cannot be used. The effect
exists mainly in the high-conductivity limit, {\it i.e.} for
sufficiently large magnetic Reynolds numbers (see Figure
\ref{pRm_dep}). The radial magnetic field, on the other hand, must
be weak enough to remain passive so that it does not dominate the
flow. Figure \ref{pB_dep} shows that the maximum field is given by
$B_{\rm eq}$ which is {\em much} higher than the mean vertical
field, which is of the order of a few gauss on the Sun.

To estimate the value of the cross helicity at the solar surface we
shall assume a density scale height of 100 km. Then one finds from
Equation (\ref{12}) that
\begin{equation}
{\langle u'_r B'_r\rangle}  \simeq -  \frac{\langle B_r\rangle}{1
{\rm G}}\frac{\eta_{12}}{H_7} \;\mbox{G km s}^{-1}. \label{15}
\end{equation}
The average is here to be applied over many turbulent cells which,
in the Sun, might correspond to 30--100 Mm. The magnetic diffusivity
in Equation (\ref{15}) has been used in the form $\eta_{\rm T}=
10^{12} \eta_{12}$\,cm$^2$\,s$^{-1}$ and the density scale height as
$H_\rho= 100 H_7$ km. We thus predict the existence of a cross
helicity of more than 1 G\,km\,s$^{-1}$. We also emphasize that the
cross helicity is anti-correlated to the mean radial magnetic field,
{\it i.e.}
\begin{equation}
\langle u'_r B'_r\rangle \   \langle B_r \rangle < 0.
\label{16}
\end{equation}
For a dipolar background field the sign of the cross helicity will
be opposite in the two hemispheres.

Relation (\ref{12}) can also be used to measure the magnetic
diffusivity if the cross helicity is known by observations. In order
to find the cross helicity one only has to correlate the observed flow
fluctuations with observed magnetic fluctuations. Together with the
calculated mean value of the radial magnetic field, Equation
(\ref{12}) provides the unknown quantity $\eta_{\rm T}$. We hope
that such an analysis of the observations using, for example, data
from the Hinode satellite will soon provide supporting evidence for
an anti-correlation between $\langle u'_r B'_r\rangle$ and $\langle
B_r \rangle$, and that a meaningful value of $\eta_{\rm T}$ can be
obtained in that way.

%% Appendix
%
% \appendix

%%%%%%%%%%%%%%%%%%%%%%%%%%%%%%%%%%%%%%%%%%%%%%%%%%%%%%%%%%%%%%%%%%%%%%%%%%%
%% Acknowledgements
%
 \begin{acks}
The authors are thankful to Hongqi Zhang for attracting attention to the cross-helicity problem. This work was supported by the Deutsche
Fo\-r\-schungs\-ge\-mein\-schaft and by the Russian Foundation for
Basic Research (projects 10-02-00148, 10-02-00391).
We acknowledge the allocation of computing resources provided by the
Swedish National Allocations Committee at the Center for
Parallel Computers at the Royal Institute of Technology in
Stockholm and the National Supercomputer Centers in Link\"oping.
This work was supported in part by
the European Research Council under the AstroDyn Research Project No.\ 227952
and the Swedish Research Council Grant No.\ 621-2007-4064.
 \end{acks}

%%% %%%%%%%%%%%%%%%%%%%%%%%%%%%%%%%%%%%%%%%%%%%%%%%%%%%%%%%%%%%
%% Bibliography
%
% Using BibTeX
%
% \bibliographystyle{spr-mp-sola}
% %\bibliographystyle{spr-mp-sola-cnd} %% Alternative style: no title, no concluding page
% \bibliography{<bib file>}
%
% Without BibTeX

\end{article}
\end{document}